\documentclass[reprint,aps,prl,superscriptaddress,showpacs]{revtex4-1}

\pdfoutput=1

\usepackage{amsmath,amssymb}
\usepackage{graphicx}
\usepackage{color}
\definecolor{darkred}{rgb}{0.55,0.00,0.00}
\definecolor{darkgreen}{rgb}{0.00,0.39,0.00}
\usepackage{units}
\usepackage{braket}
\usepackage[super]{nth}

\renewcommand{\citet}{\cite} 
\renewcommand{\citep}{\cite}
\graphicspath{{graphics/}} 

\renewcommand{\figurename}{Fig.}
\makeatletter
\renewcommand{\fnum@figure}{\textbf{Figure~\thefigure}}
\makeatother
\newcommand{\suba}{\textbf{a}}\newcommand{\refa}{a}
\newcommand{\subb}{\textbf{b}}\newcommand{\refb}{b}
\newcommand{\subc}{\textbf{c}}\newcommand{\refc}{c}
\newcommand{\subd}{\textbf{d}}\newcommand{\refd}{d}
\newcommand{\sube}{\textbf{e}}\newcommand{\refe}{e}
\newcommand{\subf}{\textbf{f}}\newcommand{\reff}{f}

\usepackage{hyperref}
\hypersetup{colorlinks=true,final=true,
        linkcolor=black,
        citecolor=black,
        filecolor=black,
        urlcolor=darkred,
        plainpages=false,
        pdfauthor={C. Lang},
        pdftitle={Probing Correlations, Indistinguishability and Entanglement in Microwave Two-Photon Interference}
}

\begin{document}

\title{Probing Correlations, Indistinguishability and Entanglement in\\
Microwave Two-Photon Interference}

\author{C.~Lang}
\author{C.~Eichler}
\author{L.~Steffen}
\author{J.~M.~Fink}
\altaffiliation{Now at Institute for Quantum Information and Matter, California Institute of Technology, Pasadena, CA 91125, USA}
\affiliation{Department of Physics, ETH Zurich, 8093 Zurich, Switzerland.}
\author{M.~J.~Woolley}
\altaffiliation{Now at School of Engineering and Information Technology, UNSW Canberra, Canberra, ACT 2600, Australia}
\author{A.~Blais}
\affiliation{D\'epartement de Physique, Universit\'e de Sherbrooke,
Sherbrooke, Qu\'ebec, J1K 2R1 Canada.}
\author{A.~Wallraff}
\affiliation{Department of Physics, ETH Zurich, 8093 Zurich, Switzerland.}
\date{January 15, 2013}


\maketitle

\textbf{Interference at a beam splitter reveals both classical and quantum properties of electromagnetic radiation. When two indistinguishable single photons impinge at the two inputs of a beam splitter they coalesce into a pair of photons appearing in either one of its two outputs. This effect is due to the bosonic nature of photons and was first experimentally observed by Hong, Ou, and Mandel (HOM)~\citep{Hong1987}. Here, we present the observation of the HOM effect with two independent single-photon sources in the microwave frequency domain. We probe the indistinguishability of single photons, created with a controllable delay, in time-resolved second-order cross- and auto-correlation function measurements. Using quadrature amplitude detection we are able to resolve different photon numbers and detect coherence in and between the output arms. This measurement scheme allows us to observe the HOM effect and, in addition, to fully characterize the two-mode entanglement of the spatially separated beam splitter output modes. Our experiments constitute a first step towards using two-photon interference at microwave frequencies for quantum communication and information processing, e.g.~for distributing entanglement between nodes of a quantum network~\citep{Kimble2008,Duan2010} and for linear optics quantum computation~\cite{Knill2001,OBrien2009}.}

Presently, HOM two-photon interference has been demonstrated exclusively using photons at optical or telecom wavelengths. Experiments were performed with photons emitted from a single source using parametric downconversion~\cite{Hong1987}, trapped ions~\cite{Duan2010}, atoms~\citep{Legero2004}, quantum dots~\citep{Santori2002} and single molecules~\citep{Kiraz2005}. The HOM effect has also been observed with two independent sources~\citep{Riedmatten2003, Kaltenbaek2006, Beugnon2006, Maunz2007, Flagg2010, Patel2010, Lettow2010} realizing indistinguishable single-photon states which are required as a resource in quantum networks or linear optics quantum computation. Such experiments have also been performed using donor impurities as sources~\cite{Sanaka2009} including NV-centers in diamond~\citep{Sipahigil2012, Bernien2012}. Furthermore, the HOM effect has been employed to create entanglement between ions~\citep{Moehring2007} in spatially separated traps, and to realize a controlled-NOT gate in a small-scale photonic network~\citep{Shadbolt2012}.

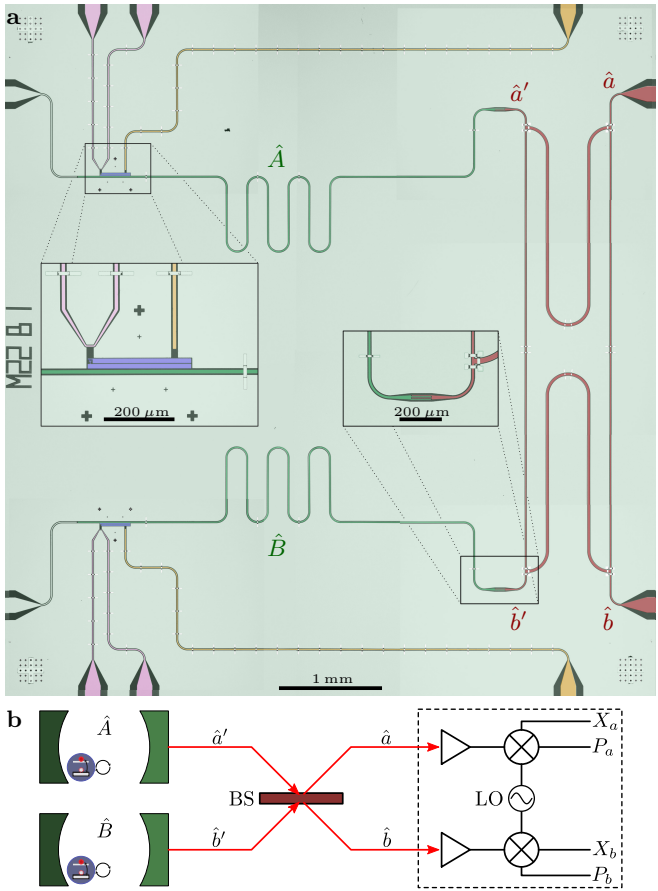
\begin{figure}[hbt]
  \includegraphics[clip,trim= 150 150 150 150,angle=90,width=246pt]{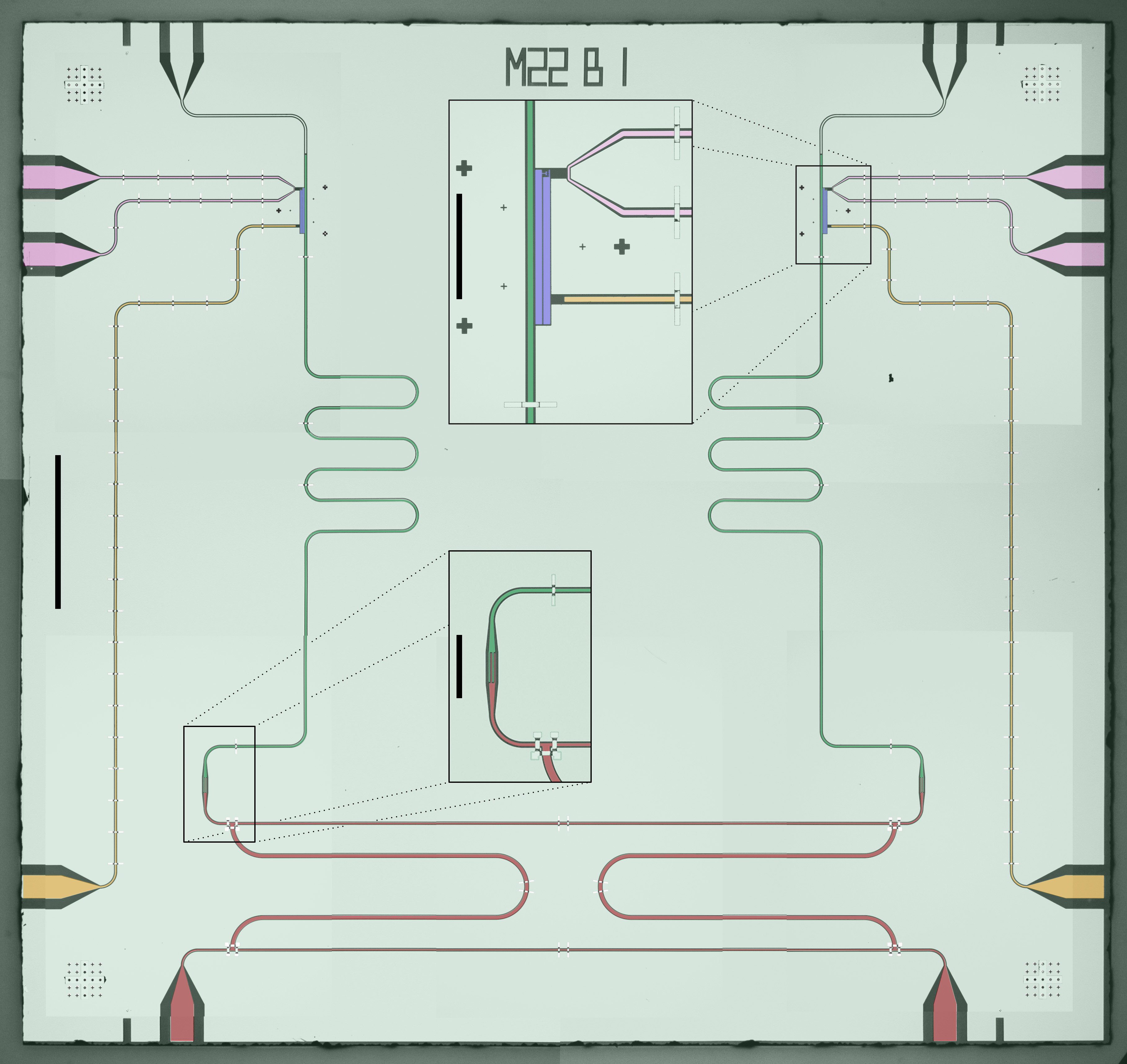}%
  \begin{picture}(0,0)%
    \put(-147, 200){\color{darkgreen}{$\hat{A}$}}%
    \put(-147,  53){\color{darkgreen}{$\hat{B}$}}%
    \put( -55, 225){\color{darkred}{$\hat{a}'$}}%
    \put( -55,  25){\color{darkred}{$\hat{b}'$}}%
    \put( -20, 230){\color{darkred}{$\hat{a}$}}%
    \put( -20,  25){\color{darkred}{$\hat{b}$}}%
    \put(-130,   5){\tiny$\unit[1]{mm}$}%
    \put(-205, 106){\tiny$\unit[200]{\mu m}$}%
    \put( -99, 106){\tiny$\unit[200]{\mu m}$}%
  \end{picture}\\[0.1cm]%
  \begin{picture}(0,0)\put(-12,328){\suba}\put(-12,62){\subb}\end{picture}%
  \includegraphics[clip,trim= 18 66 121 66,width=221pt]{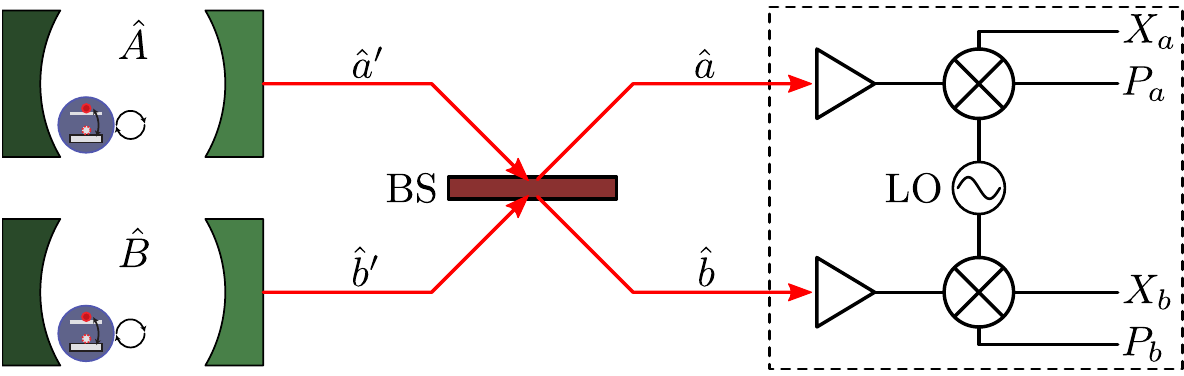}%
  \caption{\textbf{Sample and schematic.} \suba,~False color micrograph of the HOM device. A transmon qubit~(blue, see inset) controlled by a flux (purple) and charge gate line (orange) through which current or microwave pulses are applied, respectively, is coupled to each coplanar waveguide resonator (green) with fundamental mode~$\hat{A}\,(\hat{B}$). The coupling rate of the input port of the asymmetric resonator is by a factor of $10^3$ smaller than at the output port. Each resonator output is coupled to the input mode~$\hat{a}'\,(\hat{b}')$~(inset) of a microwave beam splitter~(dark red) with two output modes $\hat{a}\,(\hat{b})$. \subb,~Schematic of the HOM experiment with two individual cavity QED systems for photon generation, a beam splitter (BS), and linear amplification and heterodyne detection for correlation function measurements and full quantum state analysis~\cite{Eichler2012}.}
  \label{fig:expSetup}
\end{figure}

Here, we demonstrate the HOM interference of two indistinguishable microwave photons emitted from independent triggered sources realized in superconducting circuits. In each source $A\,(B)$ a transmon-type qubit is strongly coupled with rate $g/2\pi = \unit[169\,(177)]{MHz}$ to a transmission line resonator, see~\figurename~\ref{fig:expSetup}\refa. To create a single photon using one of the sources, we coherently excite the qubit into a state $\alpha \ket{g} + \beta \ket{e}$ using a resonant microwave pulse. Then, we swap the qubit state into the resonator mode $\hat{A}\,(\hat{B})$ by tuning the qubit transition frequency for half a vacuum-Rabi period into resonance with the resonator. This creates the state $\alpha \ket{0} + \beta \ket{1}$ in the resonator, which for $\beta = 1$ corresponds to a single-photon Fock state~\citet{Bozyigit2011}. The photon then decays exponentially with Lorentzian spectrum through the strongly coupled output port of the resonator into the input mode $\hat{a}'\,(\hat{b}')$ of the beam splitter at a rate $\kappa/2\pi = \unit[4.1\,(4.6)]{MHz}$, see \figurename~\ref{fig:expSetup}. The two photons then interfere at the beam splitter and are emitted into the output modes $\hat{a}$ and $\hat{b}$, see \figurename~\ref{fig:expSetup}\refa. Using the dispersive interaction between qubit and resonator, we tune the emission frequencies of the two sources to an identical value of $\nu_r = \unit[7.2506]{GHz}$. This is achieved by adjusting the qubit transition frequencies~$\nu_a$ to $\unit[8.575\,(8.970)]{GHz}$ with a static magnetic flux applied to the SQUID-loop of each qubit. For our experiments, we sequentially create $20$ single photons in each source at a rate $1/t_r = 1/\unit[512]{ns}\sim\unit[1.95]{MHz}$ in a sequence repeated every $\unit[12.5]{\mu s}$.

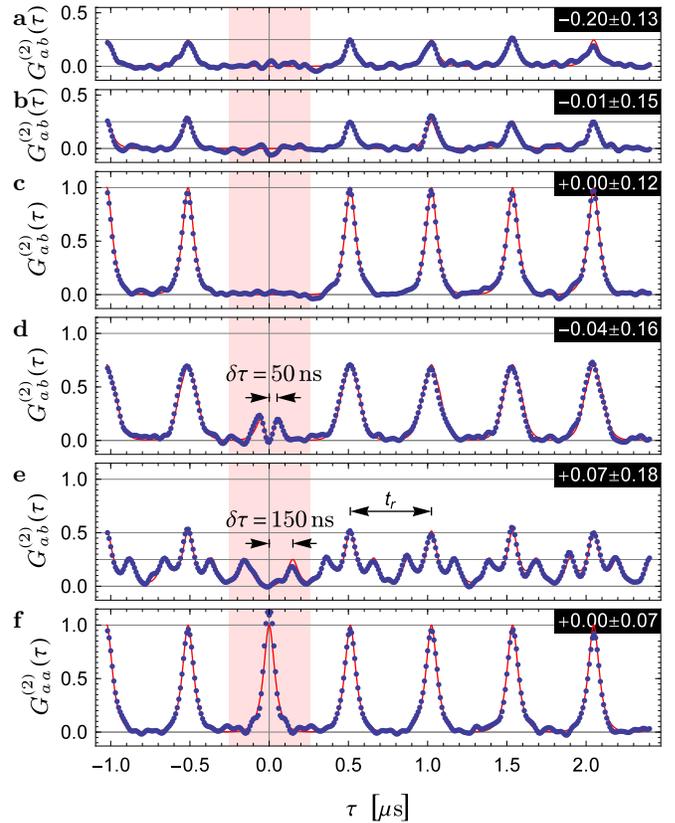
\begin{figure}
  \includegraphics[clip,trim=0 28 0 0]{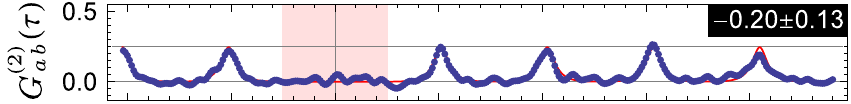}%
    \begin{picture}(0,0)\put(-246, 22){\suba}\end{picture}\\%
  \includegraphics[clip,trim=0 28 0 0]{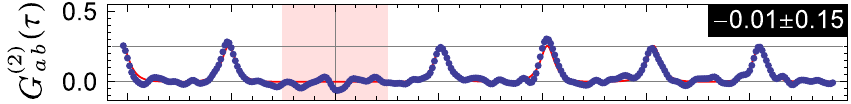}%
    \begin{picture}(0,0)\put(-246, 22){\subb}\end{picture}\\%
  \includegraphics[clip,trim=0 28 0 0]{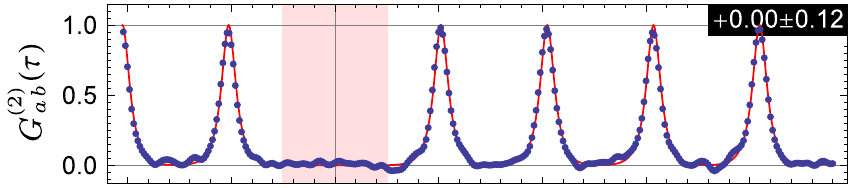}%
    \begin{picture}(0,0)\put(-246, 46){\subc}\end{picture}\\%
  \includegraphics[clip,trim=0 28 0 0]{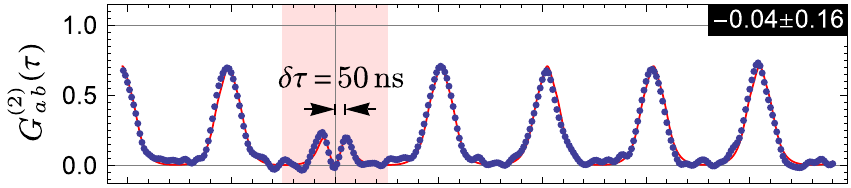}%
    \begin{picture}(0,0)\put(-246, 46){\subd}\end{picture}\\%
  \includegraphics[clip,trim=0 28 0 0]{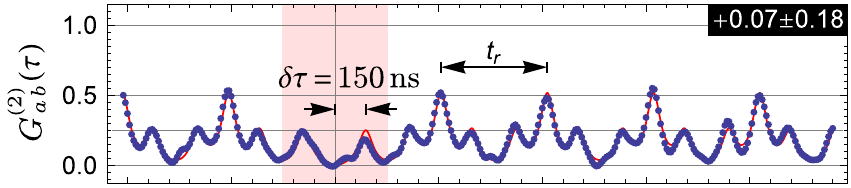}%
    \begin{picture}(0,0)\put(-246, 46){\sube}\end{picture}\\%
  \includegraphics[clip,trim=0  0 0 0]{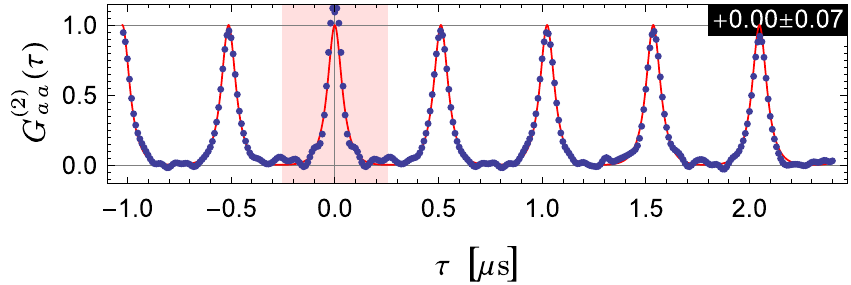}%
    \begin{picture}(0,0)\put(-246, 74){\subf}\end{picture}%
  \caption{\textbf{Second-order correlation function measurements.} \emph{Cross}-correlation $G^{(2)}_{ab}(\tau)$ of individual single photon sources displaying single photon antibunching, source $A$ in panel~\suba\ and $B$ in panel~\subb. \suba-\sube,~$G^{(2)}_{ab}(\tau)$ when operating both sources simultaneously with delay times $\delta\tau = \unit[0, 50, 150]{ns}$ displaying the HOM effect and its dependence on the temporal photon overlap at the beam splitter. \subf,~Second-order \emph{auto}-correlation $G^{(2)}_{aa}(\tau)$ of mode~$\hat{a}$ for $\delta\tau = 0$ displaying two-photon coalescence. See main text for details and a discussion of the offsets indicated in the upper right corner of each panel. Blue dots are data and red lines are theory.}
  \label{fig:HOMG2}
\end{figure}

To probe the photon statistics in the beam splitter output modes~$\hat{a}$ and $\hat{b}$ we use two spatially separated heterodyne detection channels~\cite{Menzel2012, Eichler2012}, see dashed box in \figurename~\ref{fig:expSetup}\refb. Each channel consists of a set of semiconductor linear amplifiers, a microwave frequency mixer for downconversion to $\unit[25]{MHz}$, followed by an analog-to-digital converter. After further digital downconversion to d.c.~and filtering, we extract the two quadrature amplitudes $X_{a/b}(t)$ and $P_{a/b}(t)$ corresponding to the complex envelope of the two time-dependent signals $S_{a/b}(t) = X_{a/b}(t) + i P_{a/b}(t)$~\cite{daSilva2010}. In contrast to many other HOM experiments in which photons in the beam splitter outputs are detected by single-photon counters, our measurement of the complex envelope is intrinsically photon-number resolving for averaged measurements and allows us to measure coherences of the electromagnetic field. The complex envelopes $S_{a/b}(t)$ are used to compute the statistics required to extract all relevant quantum correlations measured here~\cite{daSilva2010,Eichler2012}. To analyze the digital data, signal processing and statistical analysis are performed in real-time using field programmable gate array (FPGA) based electronics~\cite{Bozyigit2011}. The noise added by the detection chain is fully characterized by measuring its statistical properties when the output modes~$\hat{a}$ and~$\hat{b}$ of the beam splitter are left in the vacuum state with both sources idle~\cite{Eichler2012,daSilva2010}. To verify the single-photon character of each source individually, we have measured their second-order cross-correlation functions $G^{(2)}_{ab}(\tau)$. These display clear antibunching~\cite{Bozyigit2011}, see \figurename~\ref{fig:HOMG2}\refa\refb.

To investigate two-photon quantum interference, we simultaneously generate two indistinguishable photons ideally realizing a two-mode entangled state $(\ket{20}+\ket{02})/\sqrt{2}$ at the beam splitter outputs. The measured cross-correlation of the beam splitter output powers is observed to vanish $G^{(2)}_{ab}(\tau) \approx 0$ for all $\tau$ between $-t_r/2$ and $t_r/2$, see colored region in \figurename~\ref{fig:HOMG2}\refc. Therefore, we conclude that both microwave frequency photons coalesce at the beam splitter. This is the HOM effect with microwave photons. In our experiments the spatio-temporal coherence of the single-photon states is governed by resonator decay alone, and shows no significant additional dephasing. This is in stark contrast to many other experiments in which decoherence resulting in random frequency differences between interfering photons causes finite correlations at $\tau \sim 0$~\cite{Legero2004,Beugnon2006,Flagg2010,Bernien2012}. At $\tau = n\,t_r$ ($n=\pm 1,\pm 2, \dots, \pm 10$), the peak at integer non-zero multiples of the photon generation period reflects the product of the power in each output $\braket{\hat{a}^\dagger\hat{a}} \braket{\hat{b}^\dagger\hat{b}}$ which we have normalized to one.

All measured second-order correlation functions are normalized by a common scaling factor and a relative gain between the two amplification channels. A remaining small offset subtracted from the reconstructed correlation functions is indicated together with its standard deviation (extracted from multiple measurements) in the upper right corner of each panel in \figurename~\ref{fig:HOMG2}. The measured correlation functions are in good agreement  with analytical calculations~\cite{Woolley2013} (solid lines in \figurename~\ref{fig:HOMG2}), taking into account the cavity decay rates $\kappa$ extracted from independent measurements and a fixed detection bandwidth of $\unit[20]{MHz}$ chosen to reject experimental noise outside the desired band.

To explore the level of indistinguishability between the two interfering photons we introduce a time delay $\delta \tau$ on the order of the photon decay time $1/\kappa \sim \unit[37]{ns}$. For $\delta\tau = \unit[50]{ns}$ the correlation function $G^{(2)}_{ab}(\tau)$ remains close to zero at $\tau = 0$, indicating the coalescence of those photons detected with vanishingly small time difference~\cite{Woolley2013,Legero2003,Legero2004}, see \figurename~\ref{fig:HOMG2}\refd. Typically this effect is difficult to observe with detectors of insufficient bandwidth or sources with significant dephasing rates~\cite{Flagg2010}. The small positive correlations observed at $\tau \sim \delta\tau$ are due to the decreased temporal overlap of the single-photon spatio-temporal mode functions at the beam splitter, i.e.~the increased distinguishability of the two photons. At $\tau \sim n\,t_r$ we observe broadened, lower amplitude correlations. All features are in agreement with theory (solid lines)~\cite{Woolley2013}.

For $\delta\tau = \unit[150]{ns}$ (and $\delta\tau = \unit[100]{ns}$, not shown) the envelopes of the two single-photon mode functions barely overlap at the beam splitter, resulting in fully distinguishable single photons. At $\tau = \pm\delta\tau$ and $\tau = n\,t_r \pm \delta\tau$, see \figurename~\ref{fig:HOMG2}\refe, we observe positive correlations with an amplitude $1/4$, which originate from single photons impinging on the beam splitter at different times and at different input arms, compare \figurename~\ref{fig:HOMG2}\refa\refb. At $\tau = n\,t_r$ the correlations between photons in the same beam splitter input arm sum up to $1/2$, as expected.

To clearly distinguish between single-photon antibunching and two-photon coalescence in time-resolved correlation function measurements, we have also measured the second-order auto-correlation function $G^{(2)}_{aa}(\tau)$ of mode $\hat{a}$. When operating only one single-photon source $G^{(2)}_{aa}(0)$ is expected to vanish. However, in the HOM configuration with $\delta\tau = 0$ we find $G^{(2)}_{aa}(0)= G^{(2)}_{aa}(n\,t_r) = G^{(2)}_{ab}(n\,t_r)=1$, see \figurename~\ref{fig:HOMG2}\reff, as there is a $50\%$ probability of detecting two photons in mode $\hat{a}$. All measurements of $G^{(2)}_{aa}(\tau)$ are in good agreement with calculations~\cite{Woolley2013}, both for $\delta\tau = 0$ (solid lines) and for $\delta\tau = \unit[100]{ns}$ (not shown). Here, it is interesting to note that a second-order auto-correlation function is rarely directly measured, because of the the lack of sufficiently fast single-photon detectors~\cite{Steudle2012}. In contrast, using our heterodyne detection scheme we are capable of measuring $G^{(2)}_{aa}(\tau)$ for multi-photon states.

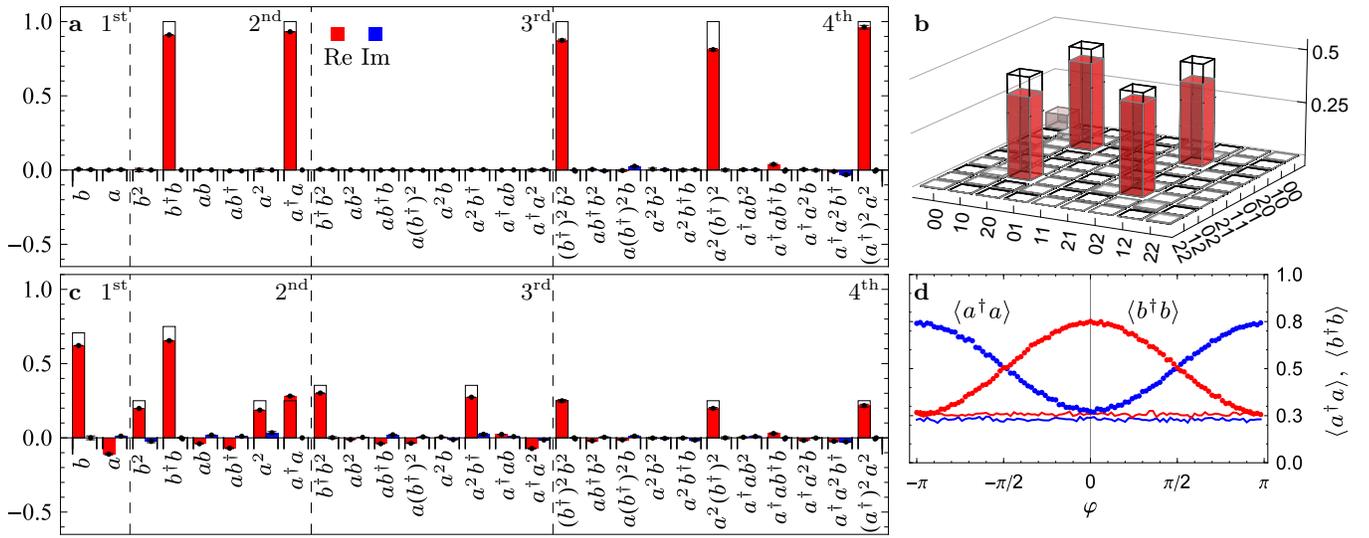
\begin{figure*}
  \includegraphics{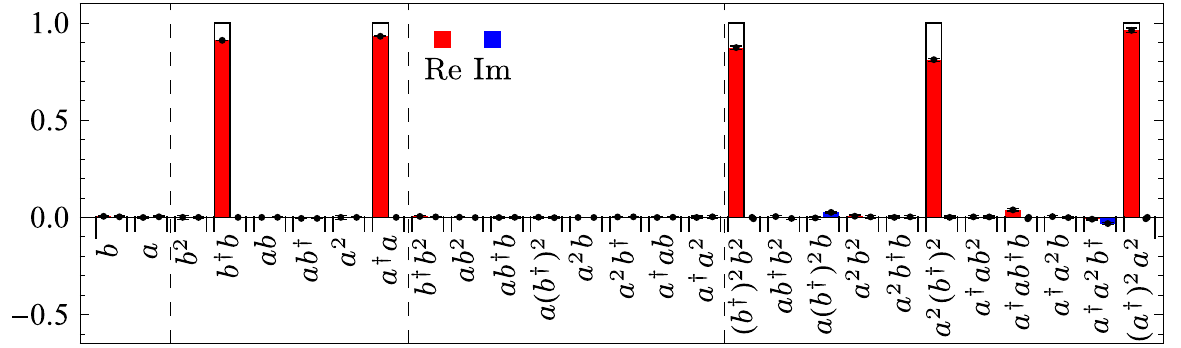}%
    \begin{picture}(0,0)\put(-315, 90){\suba}\end{picture}%
    \begin{picture}(0,0)%
      \put(-303, 90){\nth{1}}%
      \put(-247, 90){\nth{2}}%
      \put(-144, 90){\nth{3}}%
      \put( -30, 90){\nth{4}}%
    \end{picture}%
  \includegraphics[trim=2 0 0 0]{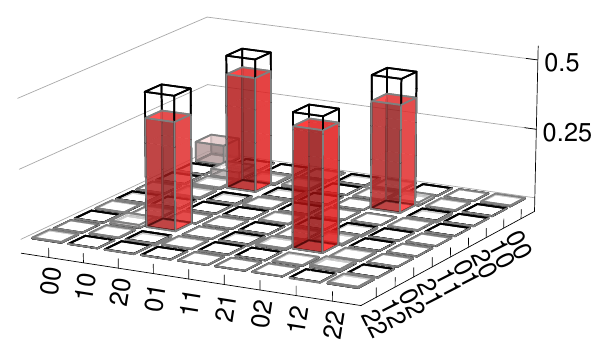}%
    \begin{picture}(0,0)\put(-164, 90){\subb}\end{picture}\\
  \includegraphics{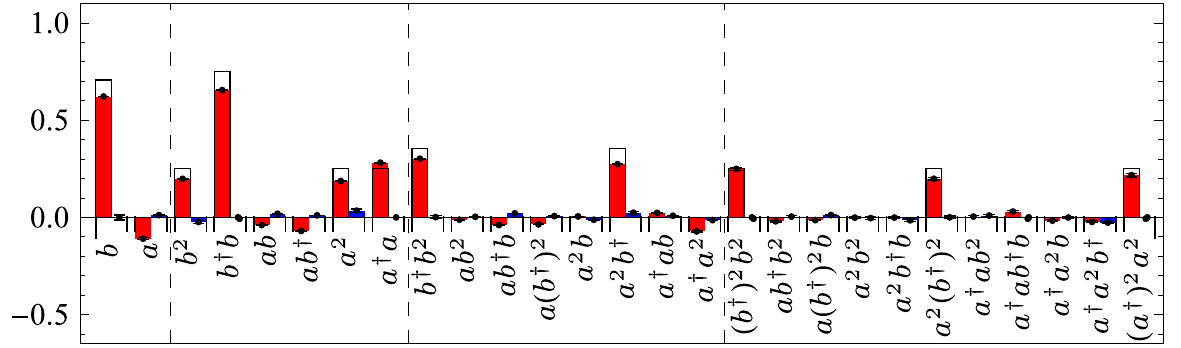}%
    \begin{picture}(0,0)\put(-315, 90){\subc}%
      \put(-303, 90){\nth{1}}%
      \put(-237, 90){\nth{2}}%
      \put(-144, 90){\nth{3}}%
      \put( -20, 90){\nth{4}}%
    \end{picture}%
  \includegraphics[trim=2 -2 0 2]{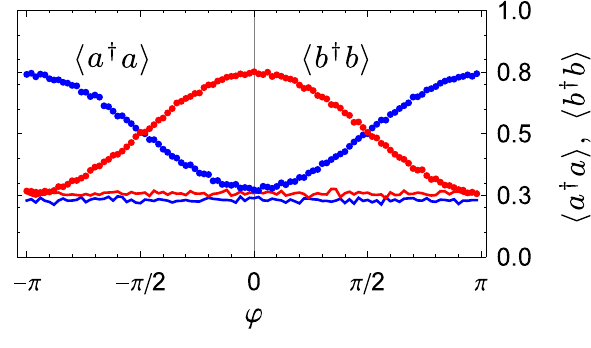}%
    \begin{picture}(0,0)\put(-164, 90){\subd}\end{picture}%
  \caption{\textbf{Full quantum state tomography of propagating photon states created in HOM experiments.}  \suba,~Measured moments $\braket{ (\hat{a}^\dagger)^n \hat{a}^m (\hat{b}^\dagger)^k \hat{b}^l }$ with $n,m,k,l \in \{0,1,2\}$ up to fourth order and \subb,~density matrix (real part) for $(\ket{20} + \ket{02})/\sqrt{2}$ with fidelity $84\%$. Colored bars are extracted from data, error bars indicate the standard deviation and wire frames represent ideal state. \subc,~Measured moments for $(\sqrt{2} \ket{00} + 2 \ket{01} + \ket{20} + \ket{02})/\sqrt{8}$ with corresponding state fidelity $84\%$. \subd,~Power detected in the two output modes ($\braket{\hat{a}^\dagger\hat{a}}$, blue) and ($\braket{\hat{b}^\dagger\hat{b}}$, red) for operating only source $B$ (line) or both sources (dots) creating superposition states with relative phase angle $\varphi$, see text for details.}
  \label{fig:HOMTom}
\end{figure*}

To distinguish between an equal mixture of the states $\ket{20}$ and $\ket{02}$, compatible with the observed correlations, and their coherent superposition $(\ket{20} + \ket{02})/\sqrt{2}$, we fully characterize the two-photon states created in our HOM experiment by quantum state tomography. This allows us to probe the entanglement generated between the coalescing two-photon states in the two output ports of the beam-splitter. The created states are also referred to as NOON states. In contrast to the NOON states of propagating photons investigated here, NOON states have also been investigated in superconducting circuits with photons localized in resonators~\cite{Wang2011b,Nguyen2012}.

To perform full quantum state tomography on propagating photons, we record four-dimensional histograms of the measured field quadratures $X_a$, $P_a$, $X_b$, and $P_b$. This is an extension of the scheme discussed in~\citet{Eichler2012} to two spatially separated modes. From these measurements we extract all moments of the two-mode field, i.e.~expectation values of the form $\braket{ (\hat{a}^\dagger)^n \hat{a}^m (\hat{b}^\dagger)^k \hat{b}^l }$ with $n,m,k,l \in \{0,1,2\}$. We observe that the first-order moments and all other odd-order moments are zero since all single-photon Fock states are characterized by a fixed photon number and consequently a fully random phase, see \figurename~\ref{fig:HOMTom}\refa. Since each mode carries exactly one photon on average, $\braket{\hat{a}^\dagger \hat{a}}$ and $\braket{\hat{b}^\dagger \hat{b}}$ are close to unity, while all other second-order moments vanish. The fourth-order moment $\braket{\hat{a}^\dagger \hat{a} \, \hat{b}^\dagger \hat{b}}$ is observed to be zero, while $\braket{\hat{a}^\dagger \hat{a}^\dagger \, \hat{a} \hat{a}}$ and $\braket{\hat{b}^\dagger \hat{b}^\dagger \, \hat{b} \hat{b}}$ are unity, consistent with the coalescence of the two photons into either output. The above observations are consistent with the ones based on the correlation function measurements in \figurename~\ref{fig:HOMG2}. Most importantly, the two-mode entanglement is indicated by the moment $\braket{ \hat{a} \hat{a} \hat{b}^\dagger \hat{b}^\dagger }$ which is close to $1$, as expected. All measured moments of the two-mode entangled state created in our HOM experiment are in good agreement with the predicted ones, see wire frames in \figurename~\ref{fig:HOMTom}\refa. Note that moments of order five and higher are all close to zero within their statistical errors.

In addition, we have determined the most likely density matrix $\rho$ characterizing the created two-mode entangled propagating photon state from the measured moments and their respective standard deviation following~\citet{Eichler2012}. We have restricted the evaluation of moments to less than three photons per output mode, since we create no more than two single photons with our sources. Note that a related analysis has recently been performed in Ref.~\citep{Israel2012}. The real part of $\rho$ is shown in \figurename~\ref{fig:HOMTom}\refb, all elements of the imaginary part of $\rho$ are smaller than $0.02$ (not shown). We extract a fidelity of the NOON type state of $\unit[84]{\%}$ and a negativity of $0.39$.

Finally, to explore the interplay between single- and two-photon interference, we have performed experiments with modes $\hat{A}$ and $\hat{B}$ prepared in superpositions of 0 and 1 photon Fock states, i.e.~$(\ket{0}-i\ket{1})/\sqrt{2}$ and  $(\ket{0}+e^{i\varphi}\ket{1})/\sqrt{2}$ with variable phase $\varphi$, ideally creating the state $(\sqrt{2} \ket{00} - i(1-e^{i \varphi}) \ket{10} + (1+e^{i \varphi}) \ket{01} + e^{i \varphi} (\ket{20} + \ket{02}))/\sqrt{8}$ at the beam splitter output. With these input states, we have first measured the power in the beam splitter output modes~$\hat{a}$ (blue line) and $\hat{b}$ (red line) when only operating source~$B$ and keeping source~$A$ idle, see \figurename~\ref{fig:HOMTom}\refd. In this case, we observe a power level corresponding to $1/4$ of a single photon independent of the phase angle $\varphi$, as expected for an equal superposition of $\ket{0}$ and $\ket{1}$ impinging on a balanced beam splitter. Operating both sources~$A$ and~$B$, we observe a sinusoidal interference with phase~$\varphi$ of the two superposition states in the beam splitter output power of mode~$\hat{a}$ (blue dots) and $\hat{b}$ (red dots), respectively. The sinusoidal oscillation is a result of the interference between one photon in either output port while the offset in power of $1/4$ is the result of two-photon coalescence.

For the phase angle $\varphi \approx 0$ we have also performed full quantum state tomography, see \figurename~\ref{fig:HOMTom}\refc. As for the two-photon NOON state we observe antibunching, coalescence, and entanglement in the moments $\braket{\hat{a}^\dagger \hat{a} \, \hat{b}^\dagger \hat{b}}$, $\braket{\hat{a}^\dagger \hat{a}^\dagger \hat{a} \hat{a}}$ and $\braket{\hat{b}^\dagger \hat{b}^\dagger \hat{b} \hat{b}}$, and $\braket{\hat{a} \hat{a} \hat{b}^\dagger \hat{b}^\dagger}$, respectively with close to expected amplitudes (wire frames). The interference of the superposition states is revealed not only in the power $\braket{\hat{a}^\dagger \hat{a}}$ and $\braket{\hat{b}^\dagger \hat{b}}$ but also in the coherences with an unbalanced number of creation and annihilation operators. The corresponding state has a fidelity of $\unit[84]{\%}$ with respect to the ideal one.

Our results suggest that multiple on-demand single-photon sources emitting indistinguishable single photons could be used for creating nonlocal entanglement in quantum repeater or quantum communication applications based on microwave photons.

\begin{acknowledgments}
This work was supported by the European Research Council (ERC) through a Starting Grant and by ETHZ. L.S. was supported by EU~IP~SOLID. A.B. and M.J.W.~were supported by NSERC, CIFAR, and the Alfred P.~Sloan Foundation.
\end{acknowledgments}

\bibliographystyle{apsrev4-1}
\bibliography{./QudevRefDB}

\end{document}